\documentclass[aps,prb,twocolumn,showpacs,superscriptaddress]{revtex4-1}

\usepackage{amsmath} 
\usepackage{graphicx} 
\usepackage{hyperref} 
\usepackage{color} 

\usepackage[caption=false]{subfig}

\newcommand{\dcaplus}{DCA$^+\:$} 

\begin{document}

\title{Two-particle correlations in a dynamic cluster approximation with continuous momentum dependence: Superconductivity in the 2D Hubbard model} 
\author{Peter Staar} \affiliation{Institute for Theoretical Physics, ETH Zurich, 8093 Zurich, Switzerland} 
\author{Thomas Maier} \affiliation{Computer Science and Mathematics Division, Oak Ridge National Laboratory, Oak Ridge TN, 37831, USA} \affiliation{Center for Nanophase Materials Sciences, Oak Ridge National Laboratory, Oak Ridge TN, 37831, USA} 
\author{Thomas C. Schulthess} \affiliation{Institute for Theoretical Physics, ETH Zurich, 8093 Zurich, Switzerland} \affiliation{Computer Science and Mathematics Division, Oak Ridge National Laboratory, Oak Ridge TN, 37831, USA} \affiliation{Swiss National Supercomputing Center, ETH Zurich, 6900 Lugano, Switzerland} 
\date{\today }
\begin{abstract}
	The \dcaplus algortihm was recently introduced in Ref.~\cite{Staar2013PRB} to extend the dynamic cluster approximation (DCA) with a continuous lattice self-energy in order to achieve better convergence with cluster size. Here we extend the \dcaplus algorithm to the calculation of two-particle correlation functions by introducing irreducible vertex functions with continuous momentum dependence consistent with the \dcaplus self-energy. This enables a significantly more controlled and reliable study of phase transitions than with the DCA. We test the new method by calculating the superconducting transition temperature $T_{c}$ in the attractive Hubbard model and show that it reproduces previous high-precision determinantal quantum Monte Carlo results. We then calculate $T_c$ in the doped repulsive Hubbard model, for which previous DCA calculations could only access the weak-coupling ($U=4t$) regime for large clusters. We show that the new algorithm provides access to much larger clusters and delivers asymptotically converged results for $T_c$ for both the weak ($U=4t$) and intermediate ($U=7t$) coupling regimes, and thereby enables the accurate determination of the exact infinite cluster size result. 
\end{abstract}


\maketitle

\section*{Introduction}

Many fascinating phenomena observed in materials, such as high-temperature superconductivity or collossal magnetoresistance, owe their existence to strong interactions between electrons and their theoretical study has therefore posed one of the most difficult challenges in condensed matter science. Due to the complexity of the underlying quantum many-body problem, analytical theories have met with limited success and numerical calculations of simplified model Hamiltonians have become increasingly important to analyze the physics of these systems. The two-dimensional (2D) Hubbard model, a standard model of correlated electron systems, has been used extensively to describe the physics of the high-temperature superconducting cuprates ~\cite{Anderson1987Science,Rice1988PRB}. Its Hamiltonian for a square lattice of sites $i$ is given by 
\begin{align}
	\label{Hubbardmodel} H = \sum_{\vec{k},\sigma} \epsilon_{\vec{k}}\, c_{\vec{k}\sigma}^{\dagger} c_{\vec{k}\sigma}^{\phantom\dagger} + U \sum_i n_{i\uparrow} n_{i\downarrow} \,. 
\end{align}
Here, $c_{\vec{k}\sigma}^{(\dagger)}$ destroys (creates) an electron with momentum $\vec{k}$ and spin $\sigma$ and $n_{i,\sigma}$ is the occupation number operator for site $i$. The dispersion 
\begin{align}
	\label{eq:dispersion} \epsilon_{\vec{k}} = -4t(\cos k_x + \cos k_y) 
\end{align}
corresponds to nearest neighbor hopping with an amplitude of $t$ and $U$ describes the on-site Coulomb repulsion between two electrons with opposite spin. 

Due to the exponential growth of the Hilbert space with the number of electrons, many numerical methods have taken a finite size approach, in which one carries out calculations on finite size lattice and then tries to scale up to the thermodynamic limit. The dynamical cluster approximation (DCA) takes a different approach in which the bulk lattice problem is replaced by an effective cluster embedded in a mean-field bath that is designed to represent the remaining degrees of freedom \cite{Hettler1998PRB,Hettler2000PRB,Maier2005RMP}. For a given cluster size, it therefore gives approximate results for the thermodynic limit and thus, in contrast to finite size methods, allows to access broken symmetry states. Similar to finite size methods, one can also carry out calculations on different cluster sizes and then use finite size scaling in order to obtain an exact result for the thermodynamic limit. 

DCA calculations on different cluster sizes have been used recently to study the normal, paramagnetic phase pseudogap state that is found in the 2D Hubbard model for electron filling factors close to one (half-filling) at intermediate to strong coupling \cite{Gull2010PRB}. Similar calculations have also shown that this model describes a superconducting transition with $d$-wave symmetry \cite{Maier2005PRL} and even allowed an analysis of the pairing interaction \cite{Maier2006PRL,Maier2006PRB}. But if one wants to carry out calculations of the doped model on large clusters at low temperatures, one has to chose an unrealistically small value of $U=4t$, since the Fermion sign problem of the QMC algorithm used as a cluster solver within the DCA prevents large cluster simulations for $U\sim 8t$ that would be more realistic for these systems. In addition, even for $U=4t$, the results for the superconducting transition temperature $T_c$ were far from converged, in part because the accessible cluster sizes were too small, but also because for small clusters, results generally depend significantly on the cluster size and shape \cite{Maier2005PRL}. 

\begin{figure}
	[t] 
	\begin{center}
		\includegraphics[width=0.5
		\textwidth]{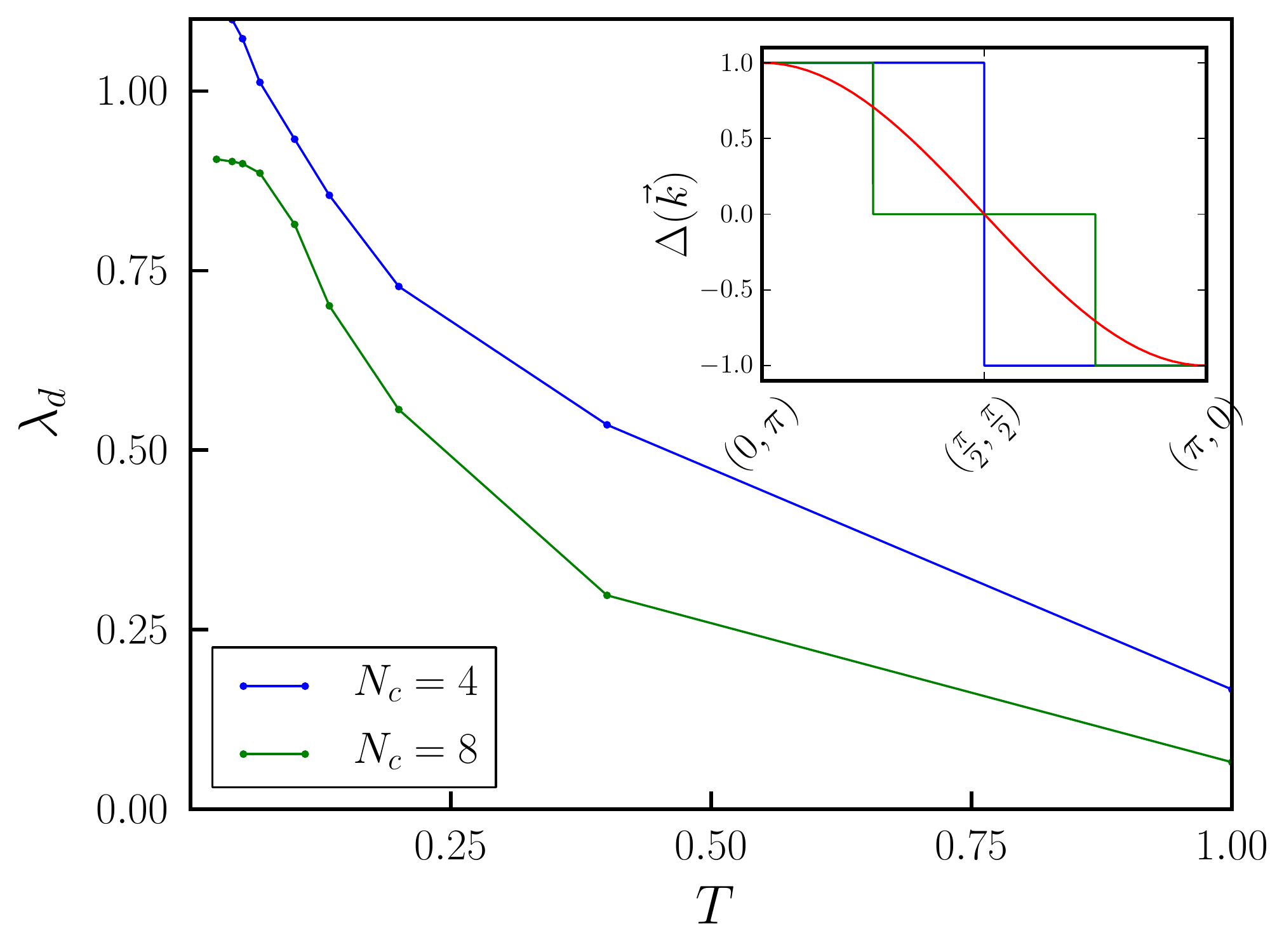} 
	\end{center}
	\caption{\label{fig:gap} DCA results for the temperature dependence of the leading $d$-wave eigenvalue $\lambda_d(T)$ of the Bethe-Salpeter equation in the particle-particle channel for the 4- and 8-site clusters calculated in a 2D Hubbard model with $U/t=8$ at 5\% doping. At $T_c$, the eigenvalue $\lambda_d(T_c)=1$. Inset: DCA approximation of the $d$-wave $\cos k_x-\cos k_y$ (red line) gap function $\Delta(k)$ in the 4- and 8-site clusters. In the 4-site cluster, $\Delta(\vec{k})$ is either 1 or -1 in the antinodal regions and misses the nodal region completely, while the 8-site cluster overestimates the nodal region. This difference in sampling of the gap function is consistent with the existence of a finite $T_c$ in the 4-site and its absence in the 8-site cluster.} 
\end{figure}

As an illustrative example of this strong cluster shape and size dependence, we plot in Fig.~\ref{fig:gap} the DCA results for the leading ($d$-wave) eigenvalue $\lambda_d$ of the Bethe-Salpeter equation in the particle-particle channel \cite{Maier2006PRB} calculated for a 2$\times$2 4-site and and 8-site cluster. This quantity is a measure of the strength of the pairing correlations in the $d$-wave channel and indicates a superconducting instability at a temperature $T_c$ where $\lambda_d(T_c)=1$. One sees that the 4-site cluster has a finite temperature superconducting transition where $\lambda_d$ crosses one, while the 8-site cluster does not. We believe that this discrepancy can be ascribed to differences in the finite size sampling of a continuous $d$-wave $\cos k_x-\cos k_y$ gap function. The red line in the inset of Fig.\ref{fig:gap} displays this function along the line from $(\pi,0)$ to $(0,\pi)$ in the first Brillouin zone. Just like the DCA self-energy, the DCA gap function is also constant within a region about the cluster $\vec{K}$ momenta, and varies between different $\vec{K}$. Since different clusters have different $\vec{K}$ points, the resulting step-function aproximation of the continuous $d$-wave gap can be very different. This is illustrated in the inset of Fig.~\ref{fig:gap} by the blue and green lines for the 4- and 8-site clusters respectively. As one sees, the 8-site cluster approximation of the gap has an extended nodal region in which the gap is zero, while the 4-site cluster approximation jumps from +1 in the region about $(0,\pi)$ to -1 near $(\pi,0)$ and the nodal region near $(\pi/2,\pi/2)$ is completely missed. This underestimation of the antinodal region in the 4-site cluster and overestimation of the nodal region in the 8-site cluster is consistent with the observed large $T_c$ in the 4-site cluster and the absence of a transition in the 8-site cluster. 

In order to reduce the DCA cluster shape and size dependence, we have recently introduced the \dcaplus algorithm, which replaces the discontinuous DCA self-energy by a continuous self-energy \cite{Staar2013PRB}. This improvement has been shown to nearly eliminate cluster shape dependencies and result in much better convergence of the self-energy as well as the pseudogap temperature with respect to the cluster size \cite{Staar2013PRB}. In addition, the \dcaplus algorithm significantly weakens the fermion sign problem and thus allows calculations on much larger clusters and interaction strengths and at lower temperatures. 

In this paper, we extend the \dcaplus framework to the two-particle level, to enable calculations of two-particle correlation functions and thus susceptibilities in order to determine possible phase transitions. In Section I, we will briefly review the DCA and \dcaplus algorithms and present the \dcaplus algorithm for calculating two-particle correlation functions with continuous momentum dependence. Then, in Sec. II, we first validate the new method by comparing \dcaplus results for the superconducting transition in the doped attractive 2D Hubbard model, for which reliable finite size QMC results on large lattices are available in the literature. Then, we discuss new \dcaplus results for the superconducting transition in the doped repulsive 2D Hubbard model for both weak and intermediate coupling regimes. 

\section{Theory and implementation}


In this section, we extend the \dcaplus algorithm that was recently introduced for the single-particle level to the two-particle level. For completeness, we first review the DCA and \dcaplus algorithms for the single-particle level and discuss the DCA formalism for the calculation of two-particle correlation functions. We then derive the \dcaplus formalism for calculating two-particle quantities from the requirement of thermodynamic consistency, which ensures that quantitites calculated from the two-particle Green's function agree with those calculated from the single-particle Green's function. Following this, we then present our algorithm for computing lattice vertex functions with continuous momentum dependence. 

\subsection{DCA and \dcaplus algorithms for single-particle correlation functions}

In the DCA \cite{Hettler1998PRB,Hettler2000PRB,Maier2005RMP}, a coarse-graining procedure is used to map the lattice problem of Eq.~(\ref{Hubbardmodel}) onto a finite size quantum impurity cluster with $N_c$ sites and periodic boundary conditions, embedded in a self-consistent mean-field. This coarse-graining procedure consists of averaging the lattice Greens-function over patches in the Brillouin zone, centered around the $N_c$ cluster-momenta $\vec{K}$. These patches are formally defined by the basis functions \cite{Okamoto2003PRB,Staar2013PRB} 
\begin{align}
	\label{patches} \phi_{\vec{K}_i}(\vec{k}) &= \left\{ 
	\begin{array}{rcl}
		1 & \forall j : \vert\vec{k}-\vec{K}_i\vert \leq \vert\vec{k}-\vec{K}_j\vert \\
		0 & \exists j : \vert\vec{k}-\vec{K}_i\vert > \vert\vec{k}-\vec{K}_j\vert 
	\end{array}
	\right. 
\end{align}
The main assumption in the DCA is that the lattice self-energy $\Sigma(\vec{k}, \varpi_m)$ is only weakly $\vec{k}$ dependent and can therefore be expanded on the patches in terms of these functions as \cite{Okamoto2003PRB} 
\begin{align}
	\label{DCASigma} \Sigma(\vec{k}, \varpi_m) &= \sum_{\vec{K}} \, \Sigma_{\vec{K}}(\varpi_m) \, \phi_{\vec{K}}(\vec{k})\,. 
\end{align}
Here, the expansion coefficients $\Sigma_{\vec{K}}(\varpi_m)$ depend only on the cluster momenta $\vec{K}$ and can therefore be calculated in the effective cluster problem. The DCA algorithm can then be summarized in a few essential steps. One starts with an initial guess for the lattice self-energy, which can be simply zero. Next, the Greens-function $G(\vec{k},\varpi_m) = (\varpi_m+\mu-\epsilon_{\vec{k}}-\Sigma(\vec{k},\varpi_m))^{-1}$ is coarse-grained over the patches, i.e. 
\begin{align}
	\label{eq:Gcg} \bar{G}_{\vec{K}}(\varpi_m) = \frac{N_c}{V}\int_V\,d\vec{k}\phi_{\vec{K}}(\vec{k}) G(\vec{k},\varpi_m) 
\end{align}
to obtain the coarse-grained Green's function $\bar{G}_{\vec{K}}(\varpi_m)$. The corresponding bare, or ''cluster-excluded'' Green's function $G^{xc}_{\vec{K}}(\varpi_m)= [\bar{G}_{\vec{K}}^{-1}(\varpi_m)+\Sigma_{\vec{K}}(\varpi_m)]^{-1}$ together with the interaction term in the Hamiltonian then defines the effective cluster problem, which, after solution, provides a new estimate for the cluster self-energy $\Sigma_{\vec{K}}(\varpi_m)$. This new estimate then provides a new parametrization of the lattice self-energy according to Eq.~(\ref{DCASigma}) in the next iteration. This process is repeated until the lattice self-energy is converged.

The expansion of $\Sigma(\vec{k},\varpi_m)$ in Eq.~(\ref{DCASigma}) in terms of the basis function $\phi_{\vec{K}}(\vec{k})$ leads to jump discontinuities between the patches. The \dcaplus algorithm uses a different approximation in order to generate a self-energy with continuous $\vec{k}$-dependence \cite{Staar2013PRB}. By multiplying Eq.~(\ref{DCASigma}) on both sides with $\phi_{\vec{K}'}(\vec{k})$ and integrating over $\vec{k}$ in the first Brillouin zone, one can effectively invert Eq.~(\ref{DCASigma}) using the orthogonality of the basis functions, i.e. $N_c/V\, \int d\vec{k} \,\phi_{\vec{K}_i}(\vec{k}) \, \phi_{\vec{K}_j}(\vec{k}) = \delta_{i,j}$, to obtain 
\begin{align}
	\label{DCAPlusSigma} \Sigma_{\vec{K}}(\varpi_m) &= \frac{N_c}{V} \int_V d\vec{k} \,\phi_{\vec{K}}(\vec{k})\,\Sigma(\vec{k},\varpi_m) \equiv \bar{\Sigma}_{\vec{K}}(\varpi_m)\,. 
\end{align}
This equation implicitly defines the lattice self-energy $\Sigma(\vec{k},\varpi_m)$ in the \dcaplus algorithm, by stating that it's coarse-grained result has to be equal to the cluster self-energy $\Sigma_{\vec{K}}(\varpi_m)$\,. The procedure to generate, given the cluster self-energy $\Sigma_{\vec{K}}(\varpi_m)$, a lattice self-energy $\Sigma(\vec{k},\varpi_m)$ with continuous and smooth $\vec{k}$-dependence is non-trivial and is typically accomplished in two consecutive steps, which involve an interpolation of $\Sigma_{\vec{K}}(\varpi_m)$ and a subsequent deconvolution of Eq.~(\ref{DCAPlusSigma}). These steps are explained in detail in Ref. \cite{Staar2013PRB}.

\subsection{Two-particle correlation functions in the DCA}

The calculation of two-particle correlation functions enables the determination of the leading correlations and possible instabilities in the system and the corresponding transition temperatures. In order to calculate these quantitites in the DCA \cite{Hettler2000PRB,Jarrell2001PRB}, one first computes the cluster one- and two-particle Green's functions (with the standard finite temperature definitions) 
\begin{align}
	\label{eq:2pG} G_{c\:\sigma}(X_1,X_2) &= - \langle T_\tau c_\sigma^{\phantom\dagger}(X_1)c_\sigma^{\dagger}(X_2)\rangle\nonumber\\
	G^{II}_{c\:\sigma_1\dots\sigma_4}(X_1,X_2;X_3,X_4) &= -\langle T_\tau c^{\phantom\dagger}_{\sigma_1}(X_1)c^{\phantom\dagger}_{\sigma_2}(X_2)\nonumber\\
	&\times c^{\dagger}_{\sigma_3}(X_3)c^{\dagger}_{\sigma_4}(X_4)\rangle\,. 
\end{align}
Here, $X_\ell=(\vec{X}_\ell,\tau_l)$ where $\vec{X}_\ell$ denotes a site in the DCA cluster and $\tau_l$ is the imaginary time, $T_\tau$ is the usual time-ordering operator, and $c^{(\dagger)}_\sigma(X)$ destroys (creates) a particle on the cluster with spin $\sigma$. Fourier-transforming on both the space and time variables gives $G_{c}(K)$ and $G^{II}_{c}(K_1,K_2;K_3,K_4)$ with $K=(\vec{K},i\omega_n,\sigma)$. Using these two quantities, one can then extract the irreducible cluster four-point vertex functions $\Gamma^{\alpha}(K_1,K_2;K_3,K_4)$. For example, in the particle-particle channel one has
\begin{align}
	\label{eq:Gamma} G^{II}_{c\,\uparrow\downarrow\downarrow\uparrow}(K_1,-K_1+Q;-K_2+Q,K_2) &=\nonumber\\
	&\hspace{-4cm}\frac{T}{N_c}\,G_{c\,\uparrow}(K_1)\,G_{c\,\downarrow}(-K_1+Q)\nonumber\\
	&\hspace{-4cm}-G_{c\,\uparrow}(K_1)\,G_{c\,\downarrow}(-K_1+Q)\,\delta_{K_1,K_2}\nonumber\\
	&\hspace{-4cm}\times\Gamma^{pp}_{c\,\uparrow\downarrow\downarrow\uparrow}(K_1,-K_1+Q;-K_3+Q,K_3)\nonumber\\
	&\hspace{-4cm}\times G^{II}_{c\,\uparrow\downarrow\downarrow\uparrow}(K_3,-K_3+Q;-K_2+Q,K_2)\,, 
\end{align}
which defines the irreducible particle-particle vertex 
\begin{align}
	\Gamma^{pp}_c(K_1,-K_1+Q,-K_2+Q,K_2) \equiv \Gamma^{pp}_{c,Q}(K_1,K_2) 
\end{align}

\noindent
for the cluster. Here, using momentum, energy and spin conservation, the dependence on 4 variables has been reduced to 3 variables with $Q=(\vec{Q},\nu)$ with the transferred momentum $\vec{Q}$ and Bosonic Matsubara frequency $\nu$. Here we have dropped the spin indices to simplify the notation for the remainder of this section. A similar expression is obtained in the particle-hole channels. Furthermore, because of the rotational invariance of the Hubbard model, it is convenient to separate the particle-particle channel into singlet and triplet and the particle-hole channel into a magnetic part which carries spin $S=1$ and a charge density part which has $S=0$.

In order to calculate the two-particle Green's function of the bulk lattice problem, the DCA approximates the lattice irreducible vertex function for channel $\alpha$, $\Gamma^{\alpha}_Q(k_1,k_2)$, with the corresponding cluster quantity $\Gamma^{\alpha}_{c,Q}(K_1,K_2)$, i.e. 
\begin{align}
	\label{eq:DCAGamma} \Gamma^\alpha_Q(k_1, k_2) = \sum_{K_1, K_2} \phi_{\vec{K}_1}(\vec{k}_1)\,\Gamma_{c,Q}^\alpha(K_1, K_2)\, \phi_{\vec{K}_2}(\vec{k}_2). 
\end{align}
The Bethe-Salpeter equation for the lattice (same as Eq.(\ref{eq:Gamma}) but with $G_c(K)$ and $\Gamma^\alpha_{c,Q}(K_1,K_2)$ replaced by their lattice counterparts $G(k)$ and $\Gamma^\alpha_{Q}(k_1,k_2)$, respectively) is then used to determine the lattice four-point correlation function $G^{II}_{Q}(k_1,k_2)$ and from a summation over $k_1$ and $k_2$ one can then determine various susceptibilities (see e.g. Ref.\cite{Jarrell2001PRB}). Here we use an alternative approach in order to determine the nature of the low energy correlations: Using $G(k)$ and $\Gamma^\alpha_Q(k_1,k_2)$, we calculate the Bethe-Salpeter eigenvalues and eigenvectors \cite{Maier2006PRL,Maier2006PRB}. For example, in the particle-particle channel with $Q=0$ 
\begin{align}
	\label{eq:BSE} -\frac{T}{N}\sum_{k_2} \Gamma^{pp}(k_1,-k_1;-k_2,k_2)\, G_\uparrow(-k_2)\,G_\downarrow(k_2) g_\alpha(k_2)\nonumber\\
	= \lambda_\alpha g_\alpha(k_1)\,. 
\end{align}
with a similar equation for the particle-hole channels. Here the sum over $k_2$ denotes a sum over both momentum $\vec{k}_2$ and Matsubara $\varpi_{2}$ variables. Instabilities of the system towards an ordered phase are signaled by an eigenvalue $\lambda_\alpha$ that crosses 1, and the momentum and frequency structure of the order parameter is reflected in the corresponding eigenvector $g_\alpha(k)$. Using the DCA approximation in Eq.~(\ref{eq:DCAGamma}) for the lattice vertex $\Gamma^{pp}$, one can then sum (coarse-grain) over the Green's function legs to obtain an equation that only depends on coarse-grained and cluster quantities \cite{Maier2006PRL,Maier2006PRB}
\begin{align}
	\label{eq:BSEcg} -\frac{T}{N_c}\sum_{K_2} \Gamma^{pp}_c(K_1,-K_1;-K_2,K_2) \bar{\chi}_0^{pp}(K_2) g_\alpha(K_2)\nonumber\\
	= \lambda_\alpha g_\alpha(K_1)\,. 
\end{align}

with

\begin{align}
	\bar{\chi}_0^{pp}(K) = \int d\vec{k}\,\phi_{K}(\vec{k}) G_\uparrow(-k)G_\downarrow(k)\,. 
\end{align}

While this reduces the complexity significantly, it also lowers the momentum resolution to the discrete set of cluster $\vec{K}$ momenta. Next we will discuss the \dcaplus extension to this formalism based on the computation of a lattice irreducible vertex $\Gamma^\alpha(k_1,k_2)$ with continuous momentum dependence in order to retain the full momentum resolution. 

\subsection{Thermodynamic consistency and the \dcaplus algorithm}\label{sectionC}

As discussed in the previous section, in order to extend the \dcaplus algorithm to the two-particle level, one needs to determine irreducible vertex functions $\Gamma^\alpha(k_1,k_2)$ for channel $\alpha$ with continuous momentum dependence given the cluster vertex functions $\Gamma^\alpha_c(K_1,K_2)$. Thermodynamic consistency in the Baym-Kadanoff sense \cite{Baym1961PR} ensures that observables calculated from the single-particle Green's function agree with those calculated from the two-particle Green's function (or equivalently as derivatives of the lattice grand potential). In this sense, the relation between $\Gamma^\alpha(k_1,k_2)$ and $\Gamma^\alpha_c(K_1,K_2)$ should be consistent with the \dcaplus relation on the single-particle level between the lattice self-energy $\Sigma(k)$ and the cluster self-energy $\Sigma_c(K)$. An algorithm is thermodynamically consistent if it is self-consistent and if the irreducible vertex functions are related to the self-energy according to
\begin{align}
	\label{eq:Gamma-Sigma} \Gamma^\alpha(k_1,k_2) &= \frac{\delta \Sigma(k_1)}{\delta G(k_2)}\,. 
\end{align}

\noindent
Here $\alpha$ denotes the channel (particle-hole, spin $S=0$ and $S=1$, or particle-particle singlet or triplet) as well as transferred momentum $Q$ of the irreducible vertex corresponding to different combinations of $k_1 = (\vec{k}_1,\varpi_{1},\sigma_1)$ and $k_2 = (\vec{k}_2,\varpi_{2},\sigma_2)$. In order to satisfy thermodynamic consistency, one therefore has to find a continuous lattice irreducible vertex function $\Gamma^\alpha(k_1,k_2)$, which is related to the continuous \dcaplus lattice self-energy through Eq.~(\ref{eq:Gamma-Sigma}). By multiplying this equation on both sides with $\phi_{\vec{K}_1}(\vec{k}_1)$ and $\phi_{\vec{K}_2}(\vec{k}_2)$, respectively, and integrating over $\vec{k}_1$ and $\vec{k}_2$, one obtains with Eq.~(\ref{DCAPlusSigma})
\begin{align}
	\label{eq:GammaL-GammaC} \int d\vec{k}_1 d\vec{k}_2 &\phi_{\vec{K}_1}(\vec{k}_1) \Gamma^\alpha(k_1,k_2) \phi_{\vec{K}_2}(\vec{k}_2) \nonumber\\
	=&\int d\vec{k}_1 d\vec{k}_2 \phi_{\vec{K}_1}(\vec{k}_1) \frac{\delta\Sigma(k_1)}{\delta G(k_2)} \phi_{\vec{K}_2}(\vec{k}_2) \nonumber\\
	=&\int d\vec{k}_2 \frac{\delta \Sigma_{K_1}}{\delta G(k_2)}\phi_{\vec{K}_2}(k_2) \nonumber\\
	=&\sum_{K_3}\int d\vec{k}_2 \frac{\delta \Sigma_{K_1}}{\delta G_{K_3}} \frac{\delta G_{K_3}}{\delta G(k_2)}\phi_{\vec{K}_2}(k_2)\,. 
\end{align}

\noindent
Then, by using the relations $\delta G_{K_3}/\delta G(k_2)=\phi_{\vec{K}_3}(\vec{k}_2)$ and $\int d\vec{k}_2 \phi_{\vec{K}_3}(\vec{k}_2) \phi_{\vec{K}_2}(\vec{k}_2) = \delta_{\vec{K_3},\vec{K}_2}$ as well as the fact that $\delta\Sigma^c_{K_1}/\delta G^c_{K_2}$ is equal to the cluster irreducible vertex $\Gamma^c_{_\alpha,K_1,K_2}$, one finds that 

\begin{align}
	\label{eq:cgVertex} \int d\vec{k}_1 d\vec{k}_2 \phi_{\vec{K}_1}(\vec{k}_1) \Gamma^\alpha(k_1,k_2) \phi_{\vec{K}_2}(\vec{k}_2)=\Gamma^\alpha_c(K_1,K_2)\,. 
\end{align}

In analogy to Eq.~(\ref{DCAPlusSigma}) for the single-particle self-energy, the \dcaplus lattice irreducible vertex function is thus related to its cluster analog through a coarse-graining relation. In the standard DCA algorithm, where $\Gamma^\alpha(k_1,k_2)$ is piecewise constant (see Eq.~(\ref{eq:DCAGamma}), this requirement is trivially satisfied, but in the \dcaplus algorithm one wants to find a $\Gamma(k_1,k_2)$ with continuous momentum dependence and without jump discontinuities that satisfies Eq.~(\ref{eq:cgVertex}). Assuming that Eq.~(\ref{eq:cgVertex}) can be inverted to determine the lattice irreducible vertex $\Gamma_\alpha(k_1,k_2)$, one can then solve the lattice Bethe-Salpeter equation in channel $\alpha$ to obtain the lattice two-particle Green's function $G^{II}_\alpha(k_1,k_2)$, or equivalently, determine the eigenvalues and eigenvectors of the matrix $\Gamma^\alpha_Q(k_1,k_2) \chi^{\alpha}_0(k_2)$. Here $\chi^{\alpha}_0(k_2) = G(k_2)G(-k_2+Q)$ in the particle-particle channel and $G(k_2)G(k_2+Q)$ in the particle-hole channels.

In the following section we will discuss a stable algorithm to solve the integral equation~(\ref{eq:cgVertex}) for $\Gamma_\alpha(k_1,k_2)$.

\subsection{Calculation of the lattice-vertex $\Gamma^\alpha(\vec{k}_1, \vec{k}_2)$}

The lattice self-energy $\Sigma(k)$ is obtained in the \dcaplus through a consecutive interpolation and deconvolution of the cluster-self-energy $\Sigma_{\vec{K}}$ \cite{Staar2013PRB}. To maintain the similarity between the vertex and the self-energy, we will follow the same procedure in order to generate an estimate of the lattice vertex functions. To simplify the interpolation-process, we first decompose the cluster vertex into its singular value representation

\begin{align}
	\Gamma^\alpha_c(K_1, K_2) &= \sum_{i} \sigma_i\, U_i(K_1)\, V_i(K_2)\,. 
\end{align}

\noindent
In this separable representation, the cluster vertex functions are written as a sum over products of functions, which depend only on a single variable ($K_1$ or $K_2$). The singular value decomposition of the cluster vertex is motivated by two reasons: First, it simplifies the interpolation of the cluster vertex, because $U_i(K_1)$ and $V_i(K_2)$ are functions of just a single $\vec{K}$ and can be interpolated independently. Second, it is often the case that the singular vectors have very strong frequency dependence, but much weaker momentum dependence. This weak momentum dependence makes them ideal functions to interpolate with cubic splines, without the risk of introducing any numerical artefacts. The interpolated vertex-function $\tilde{\Gamma}$ can thus be written as

\begin{align}
	\tilde{\Gamma}_c^\alpha(k_1, k_2) &= \sum_{i} \sigma_i\, U_i(k_1)\, V_i(k_2)\,, 
\end{align}

\noindent
where $U_i(k_1)$ and $V_i(k_2)$ with $k_i=(\vec{k}_i,\varpi_{i})$ are cubic spline interpolations in momentum space of $U_i(K_1)$ and $V_i(K_2)$, respectively, with $K_i=(\vec{K}_i,\varpi_{i})$. In the following we drop the frequency arguments for simplicity.

Just as for the self-energy, we then generalize the coarse-graining in Eq.~(\ref{eq:cgVertex}) to a convolution and expand the lattice vertex function into the same set of basis-functions $\{\mathcal{B}\}$ that is used for the lattice self-energy~\footnote{The set of basis-functions $\{\mathcal{B}\}$ can be freely chosen, since the \dcaplus is not dependent on the choice of the basis-functions. In this paper, we have used cubic Hermite splines\cite{Keys1981}.}. If cubic Hermite splines\cite{Keys1981} are used as basis-functions, the continuous lattice vertex function can be expanded as follows
\begin{align}\label{eq:Gamma_expansion}
\Gamma^\alpha(k, k') &= \sum_{i,j} \mathcal{B}_\varpi(\vec{k}-\vec{k}_i)  \gamma^\alpha(k_i, k_j)  \mathcal{B}_\varpi'(\vec{k'}-\vec{k}_j)\,.
\end{align}
Here, the vectors $k_i$ span a fine rectangular grid that covers the whole Brillouin zone. Using the explicit expansion in Eq.~(\ref{eq:Gamma_expansion}), one can rewrite Eq.~(\ref{eq:cgVertex}) as a matrix-equation,
\begin{align}
	\label{lattice_deconv} \tilde{\Gamma}^\alpha_c(k, k') &= \sum_{ \vec{k}_1, \vec{k}_2}\Phi(k, \vec{k}_1)\,\gamma^\alpha(k_1, k_2)\,\Phi(k',\vec{k}_2) \\
	\Phi_\varpi(\vec{k}_1, \vec{k}_2) &= \int d\vec{k}\, \phi_{\vec{0}}(\vec{k}_1-\vec{k})\, \mathcal{B}_\varpi(\vec{k}-\vec{k}_2) \nonumber 
\end{align}

Using a singular value decomposition of the matrix $\Phi$,
\begin{align}
	\Phi_\varpi &= \sum_{i} \sigma^{\Phi}_i\, u^{\Phi}_i(\vec{k}_1)\, v^{\Phi}_i(\vec{k}_2)\,, 
\end{align}
(note that all quantities on the right carry an implicit $\varpi$-dependence)
we can formally invert Eq.~(\ref{lattice_deconv}) and obtain an explicit formula for the lattice-vertex $\Gamma$
\begin{align}
	\label{lattice_deconv2} \Gamma^\alpha(k, k') &= \sum_{i} \sigma_i\, \tilde{u}_i(k)\, \tilde{v}_i(k'), \\
	\tilde{u}_i(\vec{k}) &= \sum_{j} \: v^{\Phi}_j(\vec{k})\, \frac{\langle u^{\Phi}_j(\vec{k}), U_i(\vec{k}) \rangle}{\sigma^{\Phi}_j} \nonumber, \\
	\tilde{v}_i(\vec{k}) &= \sum_{j} \: \frac{\langle V_i(\vec{k}), v^{\Phi}_j(\vec{k}) \rangle}{\sigma^{\Phi}_j} \: u^{\Phi}_j(\vec{k})\, \nonumber. 
\end{align}

Here, $\langle a, b \rangle$ represents the usual dot-product between the vectors $a$ and $b$. Similar to what was seen on the single-particle level in Ref.~\cite{Staar2013PRB} (see Figs. 12 and 14), we generally find that the singular values of the $\tilde{\Phi}$-matrix decay rapidly. Just as in the case of the self-energy, the lattice mapping for the vertex can only be performed if the expansion coefficients $\langle u^{\Phi}_j(\vec{k}), U_i(\vec{k}) \rangle$ and $\langle V_i(\vec{k}), v^{\Phi}_j(\vec{k}) \rangle$ decay faster than the singular values. For numerical reasons, we generally impose an upper bound to the inverse of singular values. Due to the numerical noise of the Monte Carlo integration, the expansion coefficients $\langle u^{\Phi}_j(\vec{k}), U_i(\vec{k}) \rangle$ and $\langle V_i(\vec{k}), v^{\Phi}_j(\vec{k}) \rangle$ will become small, but never really zero, As a consequence, we convert $1/\sigma^\phi$ to the value $\min\{1/\epsilon, 1/\sigma^\phi\}$, where $\epsilon$ is a small number. In this way, we take all components into account but assure that they don't diverge due to numerical inconsistencies.

\section{Phase transitions in the 2D attractive and repulsive Hubbard models}

The \dcaplus\:algorithm was previously applied \cite{Staar2013PRB} to study the self-energy and pseudogap temperature in the doped 2D repulsive Hubbard model. Here we use the new \dcaplus two-particle formalism discussed in the previous section to determine transition temperatures in both the attractive and repulsive Hubbard models. 

In order to validate the \dcaplus two-particle framework, we will study the attractive Hubbard model and calculate the superconducting (s-wave) transition temperature for various electron densities. When doped away from half-filling, this model has a finite temperature Kosterlitz-Thouless (KT) superconducting transition with a singlet s-wave order parameter (see e.g. Ref.~\cite{Scalettar1989PRL}. This model does not have a fermion sign problem, so that accurate results for the KT transition temperature $T_{\rm KT}$ have been obtained from large cluster QMC calculations\cite{Paiva2004PRB}, which we will use to validate the new algorithm. 

The 2D repulsive Hubbard model has been investigated extensively because of its relevance to the cuprate high-temperature superconductors, but the minus sign problem of the doped model has made it difficult to address the question of whether this model supports a d-wave superconducting state at high temperatures. Variational Monte Carlo (VMC) studies \cite{Imada2007JPSJ} tend to find a superconducting phase only for couplings $U\gtrsim 6$, while calculations based on cluster dynamic mean field theory generally find a transition to a superconducting state also for the weak-coupling $U\sim 4t$ regime\cite{Maier2005PRL}. For this interaction strength, previous DCA calculations  \cite{Maier2005PRL} at a filling of $\langle n\rangle =0.9$ have found a transition at $T_c \approx 0.023 t$. But the largest cluster that could be reached for these parameters had only 26 sites and the results were not converged due to the notorious cluster shape dependence of results computed with the standard DCA. Here, in light of the discrepancy with the VMC results, we will re-investigate this parameter regime using the \dcaplus algorithm. In particular we will show that its reduced minus sign problem and cluster shape dependence allows us to reach a regime with asymptotic convergence, in which the results for $T_c$ can be fitted with the expected Kosterlitz-Thouless behavior. 

Finally, we will discuss \dcaplus calculations for an intermediate coupling strength of $U=7t$, which is relevant for the cuprates. First, we will study the half-filled 2D model which has an antiferromagnetic ground state at $T=0$ but is paramagnetic at $T>0$ because of the Mermin-Wagner theorem. Due to the mean-field character of the DCA and \dcaplus, these techniques predict a finite temperature transition. We will show, however, that the transition temperature computed with \dcaplus decreases logarithmically with linear cluster size, consistent with the Mermin-Wagner theorem as seen before with DCA calculations \cite{Maier2005PRL}. Then we will study the 10\% doped model, for which previous DCA calculations could not reach $T_c$ for clusters larger than 12 sites. We will show that the \dcaplus algorithm allows us to access $T_c$ in clusters as large as 28 sites, for which asymptotic convergence is reached and $T_c$ can be reliably predicted.

\subsection{2D attractive Hubbard model}\label{attrac_Hubbard}

\begin{figure}
	[!] 
	\begin{center}
		\includegraphics[width=0.5
		\textwidth]{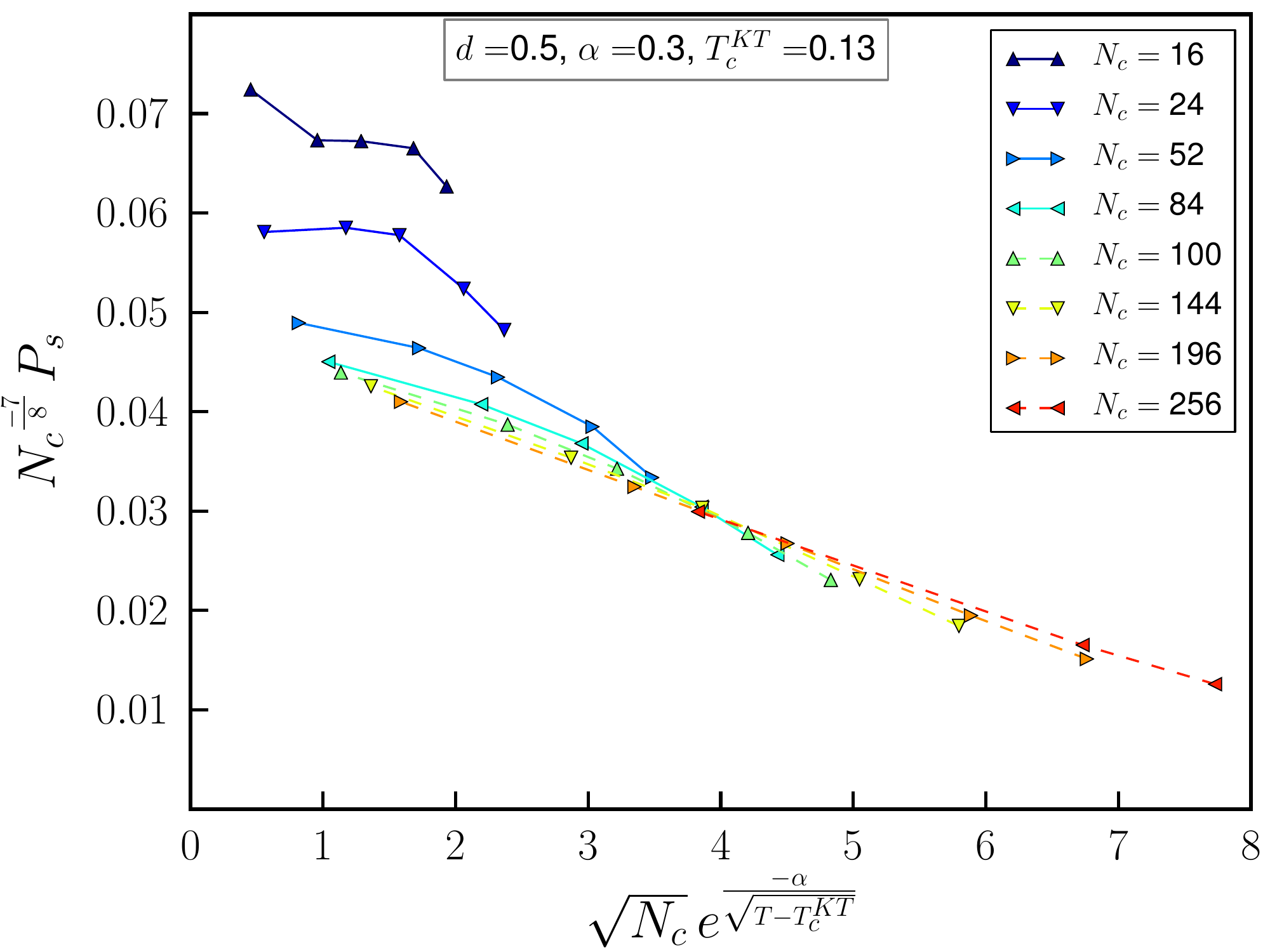} 
	\end{center}
	\caption{\label{fig:KT_scaling} Data-collapse of the cluster-susceptibility $P_s$ using the Kosterlitz-Thouless scaling form in Eq.~(\ref{PsKtScaling}) for a filling of $\langle n\rangle=0.5$. We can observe a clear data-collapse for clusters larger than 84 sites.} 
\end{figure}

The attractive Hubbard model has been studied extensively\cite{Scalettar1989PRL,Micnas1990RMP,Moreo1991PRL,Wilson2001JPCM,Paiva2004PRB} over the past three decades. Following the discovery of the high-temperature cuprates, this nontrivial toy-model has been used to shed light on the formation of Cooper pairs and other exotic states of matter which arise from the correlation between electrons. As this model does not suffer from a fermionic sign problem, large clusters can be accessed with QMC and the phase diagram can be obtained accurately through a finite size scaling procedure. The aim in this section is to validate the \dcaplus framework by reproducing the temperature versus doping phase-diagram of the attractive Hubbard model with an interaction of $U=-4$. This model has been studied in detail by Paiva et. al.\cite{Paiva2004PRB}, using finite size determinantal QMC calculations \cite{BSS1981PRD,BSS1981PRB} of large clusters for which accurate results for $T_c$ were obtained. 

\begin{figure}
	[!] 
	\begin{center}
		\includegraphics[width=0.5
		\textwidth]{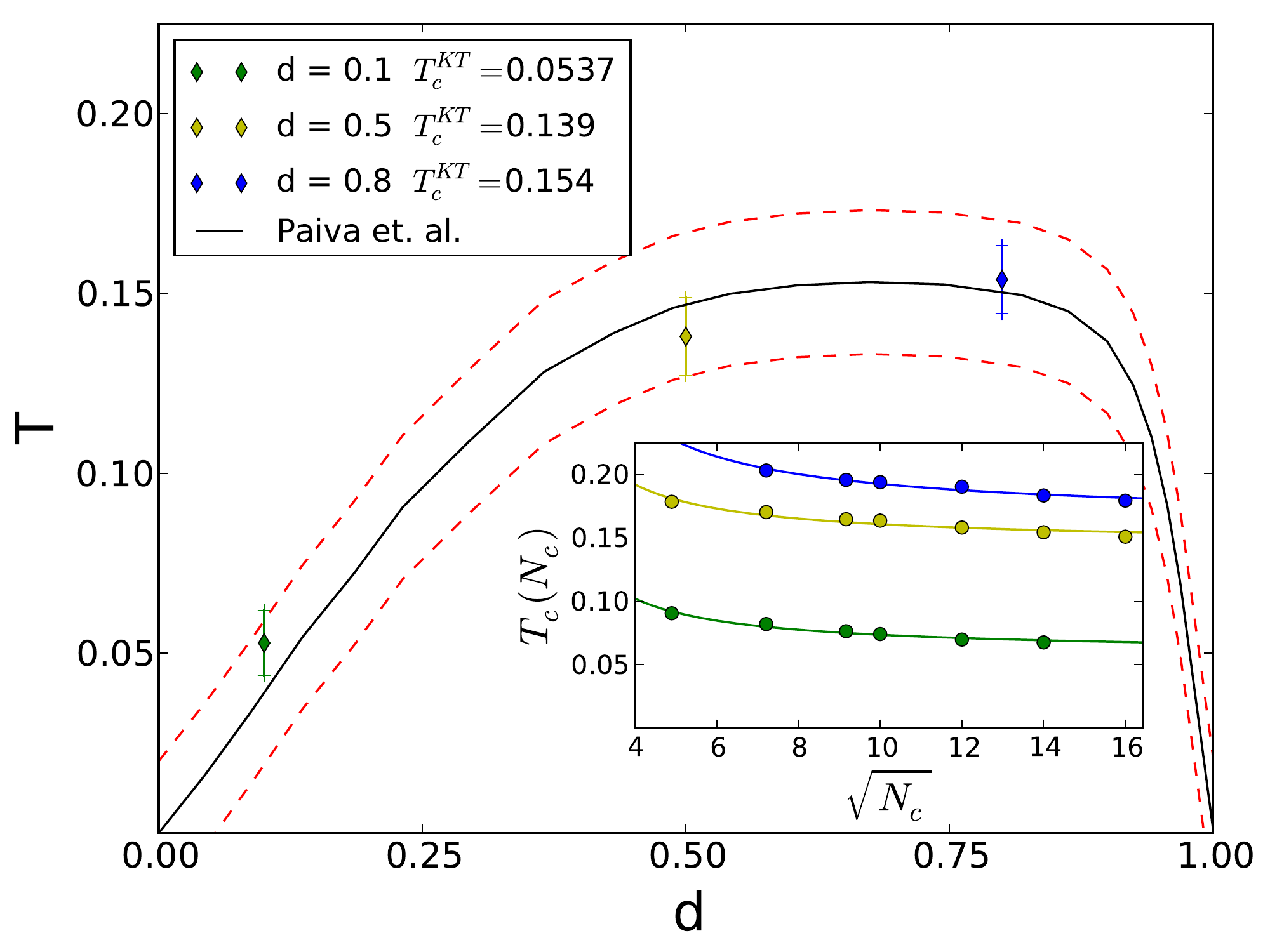} 
	\end{center}
	\caption{\label{fig:Tc_versus_d} The phase-diagram of the attractive Hubbard model with $U=-4$. The \dcaplus results lie within the error-bars (red-dotted lines) of previously reported values by Paiva et al.} 
\end{figure}
We will use two complementary procedures to determine the exact (infinite cluster size) KT transition temperature $T_{\rm KT}$: (1) We will use the same finite size scaling analysis of the cluster s-wave pair-field susceptibility that was used in Ref.~\cite{Paiva2004PRB}. This procedure avoids the determination of the lattice vertex function trough interpolation and deconvolution of the cluster vertex function. (2) We will determine the superconducting transition temperature $T_c(N_c)$ for a given cluster size $N_c$ by calculating the leading eigenvalue of the lattice Bethe-Salpeter equation in Eq.~(\ref{eq:BSE}) as outlined in Section~(\ref{sectionC}) and then obtain an estimate for $T_{\rm KT}$ by fitting $T_c(N_c)$ with the expected KT form. We will show that both procedures result in the same estimate for $T_{\rm KT}$.

We start with a finite size scaling analysis of the s-wave cluster pair-field susceptibility 
\begin{align}
	\label{clusterPs} P_s = \int_0^\beta d\tau \,\langle \Delta^\dagger(\tau)\Delta(0)\rangle 
\end{align}
with 
\begin{align}
	\label{Deltas} \Delta^\dagger = \frac{1}{\sqrt{N_c}}\sum_{\vec{K}} c^\dagger_{\vec{K}\uparrow}c^\dagger_{-\vec{K}\downarrow}\,. 
\end{align}
Note that $P_s$ can be obtained directly from the $Q=0$ cluster two-particle Green's function in the particle-particle channel, $G^{II}_{c\,\uparrow\downarrow\downarrow\uparrow}(K,K')$ (see Eq.~(\ref{eq:2pG})), as
\begin{align}
	\label{PsGII} P_s = \frac{T^2}{N^2_c} \sum_{K,K'} G^{II}_{c\,\uparrow\downarrow\downarrow\uparrow}(K,K') 
\end{align}

\noindent
where the sum over $K$ (and $K'$) implicitly contains a sum over momenta $\vec{K}$ and Matsubara frequencies $\varpi$. 

If one assumes that the transition to the superconducting phase takes place when the correlation length reaches the linear cluster size $L_c = \sqrt{N_c}$, one expects from finite size scaling for a Kosterlitz-Thouless transition that \cite{Paiva2004PRB}
\begin{align}
	\label{PsKtScaling} P_s L_c^{-7/4} = L_c \exp\left[\frac{-\alpha}{\sqrt{T-T_c}}\right]\,. 
\end{align}

In Fig.~\ref{fig:KT_scaling}, we have plotted the best data-collapse for this equation at 50\% doping. The critical temperature $T_{\rm KT}=0.13$ obtained by the data-collapse is equal to the value obtained by Paiva et. al. We believe that the discrepancy on the parameter $\alpha$ (0.3 versus 0.1) can most likely be attributed to the mean-field character of the \dcaplus algorithm.

Next, we use the new \dcaplus two-particle formalism described in Section \ref{sectionC} to calculate the lattice irreducible vertex in the particle-particle channel, $\Gamma^{pp}(k,k')$, with continuous momentum dependence. We then compute the leading eigenvalue $\lambda_s(T)$ (the corresponding eigenvector has s-wave symmetry) of the pairing matrix $\Gamma^{pp} \chi^0$ that enters the lattice Bethe-Salpeter equation (see Eq.~(\ref{eq:BSE}))). This allows us to determine the transition temperature $T_c(N_c)$ for a given cluster size $N_c$ from $\lambda_s(T_c(N_c))=1$. The exact infinite size cluster result $T_c(N_c \rightarrow \infty) \equiv T_{\rm KT}$ is then obtained from fitting the $T_c(N_c)$ data with the expected KT behavior\cite{Maier2005PRL}
\begin{align}
	\label{KT-scaling} T_c(N_c) = T_c^{\rm{KT}} + \frac{A}{[B+\,\log(\sqrt{N_c})]^2}\,. 
\end{align}

As one sees from the inset of Fig.~\ref{fig:Tc_versus_d}, the fits of the data for electron densities $\langle n \rangle=0.1$, $0.5$ and $0.8$ with the form in Eq.~(\ref{KT-scaling}) are excellent. The resulting estimates for $T_{\rm KT}(\langle n\rangle)$ are shown as symbols in the main figure. The error bars are obtained by omitting each data-point once in the corresponding $T_c(N_c)$ curves, which results in 6 different estimates for $T_{\rm KT}$ for each density and thus the standard deviation represented by the error bars. One sees that the obtained transition temperatures lie within the error-bars of Paiva et. al (red dashed lines in Fig.~\ref{fig:KT_scaling}).

From these results we can draw two important conclusions: First, the transition temperature we obtain from the data-collapse of the cluster-susceptibility is in excellent agreement with the transition temperature obtained from the lattice Bethe-Salpeter equation. The first procedure is based entirely on the two-particle cluster Greens function and thus does not involve the new procedure for determining the lattice irreducible vertex, while the second method uses the new \dcaplus two-particle framework (inversion of Eq.~(\ref{eq:cgVertex}) for the lattice vertex. This provides evidence that the algorithm we use to invert the coarse-graining of the lattice vertex in Eq.~(\ref{eq:cgVertex}) provides accurate estimates of transition temperatures for a given cluster size $N_c$, which lead to the same inifite cluster size limit as the results obtained from finite size scaling of the cluster susceptibility. Second, the \dcaplus calculations reproduce the temperature versus doping phase-diagram of the attractive Hubbard model with an interaction of $U/t=-4$ previously determined by Paiva et al. From this we conclude that the \dcaplus algorithm provides a reliable way to accurately determine phase transition temperatures. 

\subsection{2D repulsive Hubbard model}

We will start the \dcaplus study of the 2D repulsive Hubbard model by re-investigating d-wave superconductivity in the weak-coupling $U=4t$ regime for which previous DCA calculations are available \cite{Maier2005PRL}. We will then move on to the intermediate-coupling $U=7t$ regime, which has been difficult to access with standard DCA. In particular, we will show results for antiferromagnetism at half-filling and d-wave superconductivity in the doped model.

\subsubsection{Superconductivity at weak coupling}

\begin{figure}
	[t] 
	\begin{center}
		\includegraphics[width=0.5
		\textwidth]{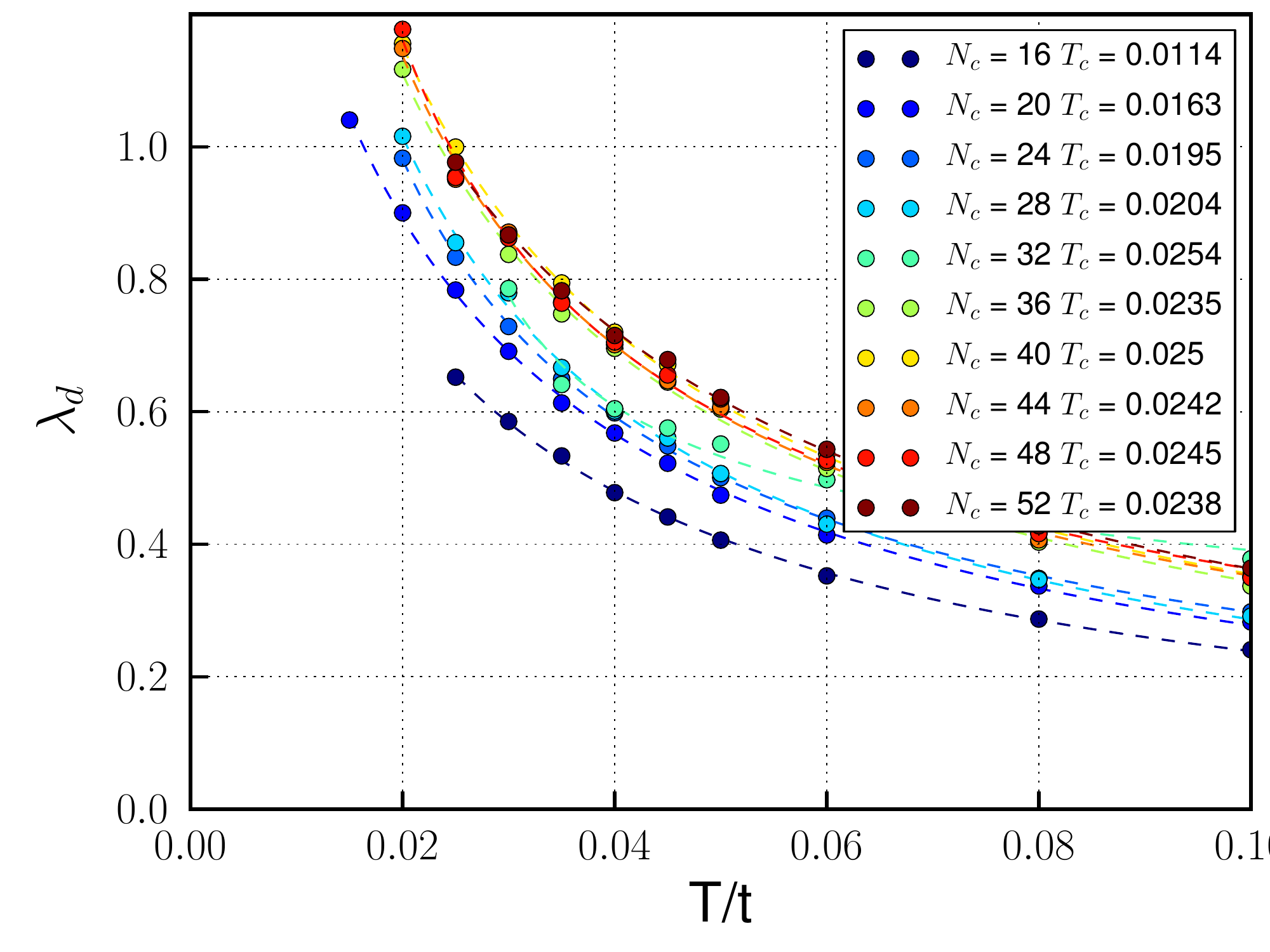} 
	\end{center}
	\caption{\label{fig:lambda_versus_T} The leading ($d$-wave) eigenvalue of the Bethe-Salpeter equation in the particle-particle channel calculated with \dcaplus in the 2D Hubbard model with $U/t=4$ and $\langle n\rangle=0.9$. } 
\end{figure}

As for the attractive model, we calculate the temperature dependence of the leading eigenvalues and eigenvectors of the pairing matrix $\Gamma^{pp}\chi^0$ that enters the lattice Bethe-Salpeter equation for different cluster sizes. At low temperatures, the leading eigenvector has d-wave symmetry. In Fig.~\ref{fig:lambda_versus_T} we show \dcaplus results for the leading d-wave eigenvalue $\lambda_d(T)$ versus temperature for cluster sizes ranging from 16 to 52 sites for $U=4t$ and $\langle n\rangle =0.9$. One sees that $\lambda_d(T)$ monotonically increases with decreasing temperature and eventually crosses one, which defines the transition temperature for a given cluster size. For the smallest cluster sizes $N_c < 36$, one also sees that at a fixed temperature, $\lambda_d$ increases monotonically with cluster size, as does $T_c$. We believe that in this regime of large $N_c$ dependence, the superconducting coherence length is larger than the cluster so that spatial phase fluctuations are neglected. Since pairs are correlated over longer distances than those within the cluster size, increasing the cluster size takes into account longer-ranged pair-field correlations and therefore $\lambda_d(T)$ and also $T_c$ increase with $N_c$. This is similar to what one sees in finite size calculations for the cluster pair-field correlations, which increase monotonically with cluster size (see e.g. Fig.~1 in Ref.~\cite{Paiva2004PRB}).

\begin{figure}
	[t] 
	\begin{center}
		\includegraphics[width=0.5
		\textwidth]{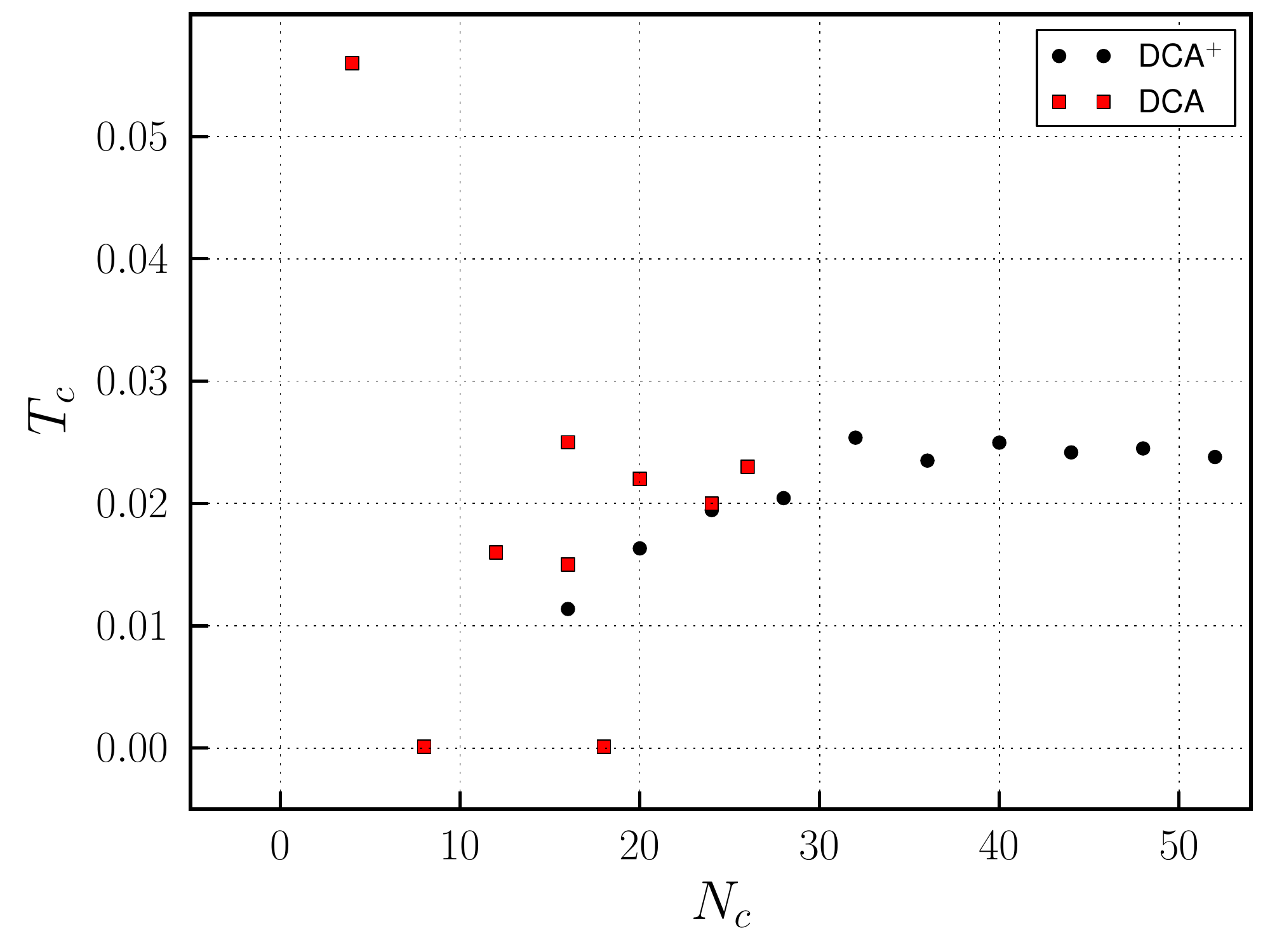} 
	\end{center}
	\caption{\label{fig:Tc_versus_Nc_4} The superconducting transition temperature $T_c$ versus cluster-size computed with DCA (red squares) and \dcaplus (black circles) in the 2D Hubbard model with $U/t=4$ and $\langle n\rangle=0.9$. The \dcaplus algorithm can access larger clusters and produces more systematic convergence.} 
\end{figure}

In order to show the $N_c$ dependence of $T_c$ more clearly, we plot in Fig.~\ref{fig:Tc_versus_Nc_4} $T_c$ versus $N_c$ as determined from $\lambda_d(T_c)=1$ (black circles) together with the previous DCA results (red squares). Here one clearly observes the monotonic rise of $T_c(N_c)$ of the \dcaplus results for $N_c < 36$. The previous DCA calculations were also able to cover most of this range in $N_c$, although the results for $T_c$ were much more erratic as can be seen from the red squares. With the new \dcaplus data, it now becomes clear that the cluster sizes that could be accessed with the DCA are in a regime where the coherence length is larger than the largest length scale covered by the clusters. The \dcaplus algorithm, however, due to the larger average QMC sign, can go to significantly larger cluster sizes. Most importantly, it can access a regime in which $T_c(N_c)$ appears to remain roughly constant with $N_c$ or just weakly decreases. We believe that in this regime, the linear cluster sizes are larger than the coherence length. In this case, just as we have found for the attractive model in Sec.~\ref{attrac_Hubbard}, $T_c$ should display a weak logarithmic decrease with cluster size according to the KT scaling behavior in Eq.~(\ref{KT-scaling}) since spatial phase fluctuations are increasingly taken into account. 

\begin{figure}
	[t] 
	\begin{center}
		\includegraphics[width=0.5
		\textwidth]{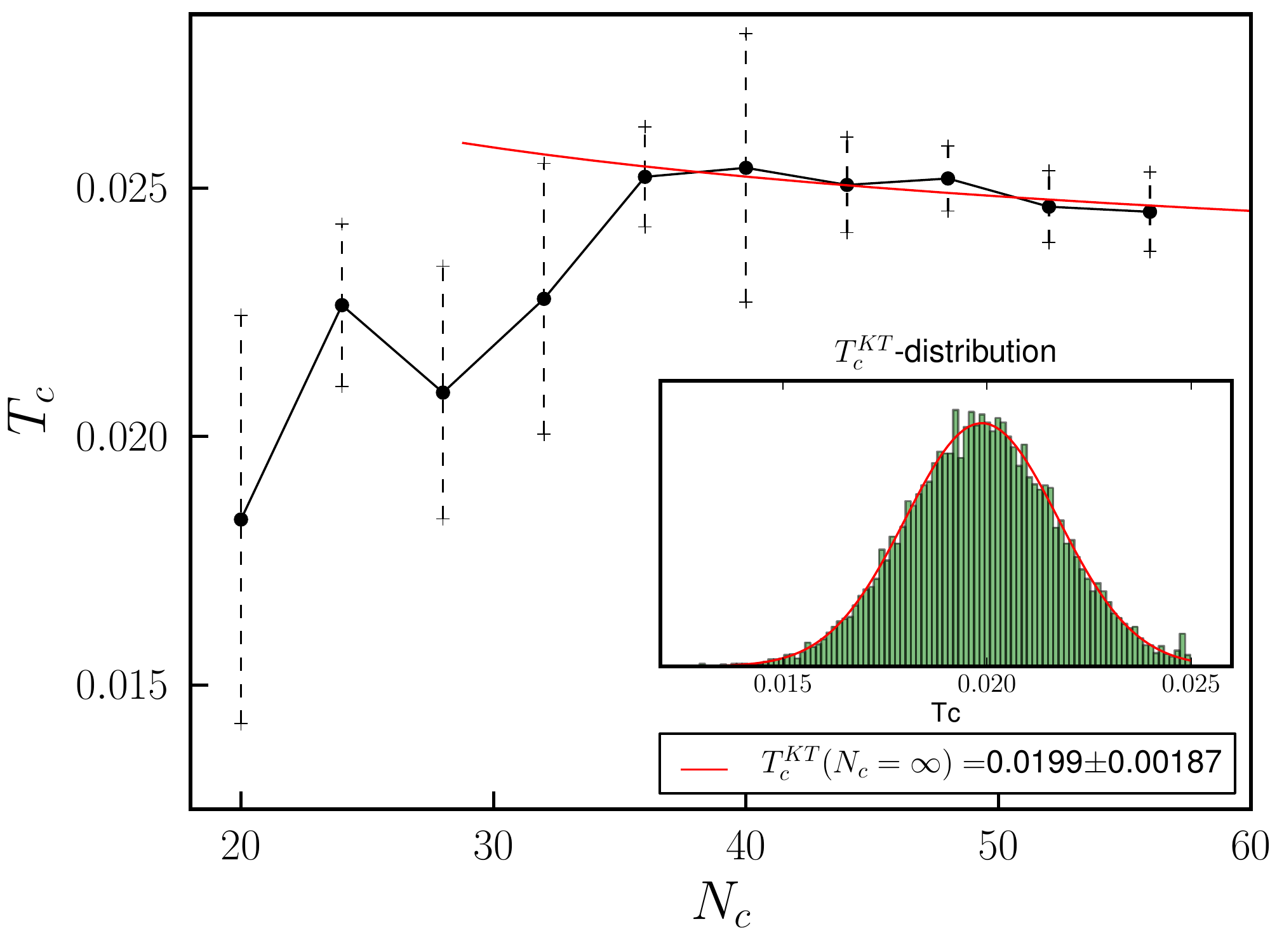} 
	\end{center}
	\caption{\label{fig:Tc_versus_Nc_4_KT} \dcaplus results for $T_c$ versus cluster size $N_c$ in the 2D Hubbard model with $U=4t$ and $\langle n\rangle = 0.9$. The symbols indicate the mean and the error bars the standard deviation of $T_c$ of four different cluster shapes with the same $N_c$. The red line shows the average fit of the Kosterlitz Thouless scaling law, Eq.~(\ref{KT-scaling}). Inset: By generating random transition temperatures for a give $N_c$, which are Gaussian distributed around the mean value and lie within the standard-deviation, we generate a distribution function for $T_c^{KT}$. This distribution is then used to obtain an estimate for the lattice transition temperature $T_c^{KT}(N_c=\infty) = 0.0199\pm 0.002$.} 
\end{figure}

Although the range of cluster sizes for which this behavior is observed is very small ($36 < N_c < 56$) and finite size scaling therefore difficult, it is interesting to see whether these results are consistent with the KT scaling behavior in Eq.~(\ref{KT-scaling}) and whether one can extract an  infinite cluster size limit $T_c(N_c\rightarrow\infty) \equiv T_{\rm KT}$. To this end we first need to determine error bars for $T_c(N_c)$. There are two sources of errors in the \dcaplus (as in the DCA) algorithm: (1) The statistical error arising from the Monte Carlo sampling, and (2) the error associated with differences in the results from different cluster shapes. While the cluster shape dependence is significantly reduced in the \dcaplus, we still assume that the statistical Monte Carlo error is smaller than the spread in results arising from different cluster shapes. Thus, for each cluster size $N_c$, we calculate $T_c$ for four different cluster shapes. The mean and standard deviation of these results is shown in Fig.~\ref{fig:Tc_versus_Nc_4_KT} as circles and dashed lines. For this calculation, we have used a very small deconvolution cut-off $\sigma_\Phi =0.1$ (typically we use $\sigma_\Phi =0.5$), which amplifies the cluster-shape dependence to a great extent. In order to obtain an estimate for $T_{\rm KT}$ and its error, we now generate for each cluster size a Gaussian distribution of 10000 transition temperatures around the mean and within the confidence interval. For each of this generated set of transition temperatures, we perform a fit with Eq.~(\ref{KT-scaling}) in order to obtain an estimate for $T_{\rm KT}$. This results in a distribution of $T_{\rm KT}$, which we show in the inset of Fig.~\ref{fig:Tc_versus_Nc_4_KT}. From a Gaussian fit of this distribution we obtain a mean of $T_{\rm KT} = 0.0199 \pm 0.0019$. The average fit to the data is shown in Fig.~\ref{fig:Tc_versus_Nc_4_KT} by the red line.

As mentioned before and demonstrated in Fig.~\ref{fig:Tc_versus_Nc_4}, the reduced cluster shape dependence of the \dcaplus in conjunction with the ability to access larger clusters allows us to identify two different regimes in the cluster size dependence of $T_c(N_c)$ separated by the superconducting coherence length $\xi$: For a linear cluster size $L_c<\xi$, $T_c(N_c)$ monotonically increases, while for $L_c>\xi$, it weakly decreases according to the KT scaling behavior. This allows us to estimate the coherence length. For the parameters in Fig.~\ref{fig:Tc_versus_Nc_4}, i.e. $U=4t$ and $\langle n \rangle =0.9$, we estimate a coherence length of $\xi \sim \sqrt{32} \approx 6$ lattice spacings.

\subsubsection{Antiferromagnetism and superconductivity at intermediate coupling}
\begin{figure}
	[t] 
	\begin{center}
		\includegraphics[width=0.5
		\textwidth]{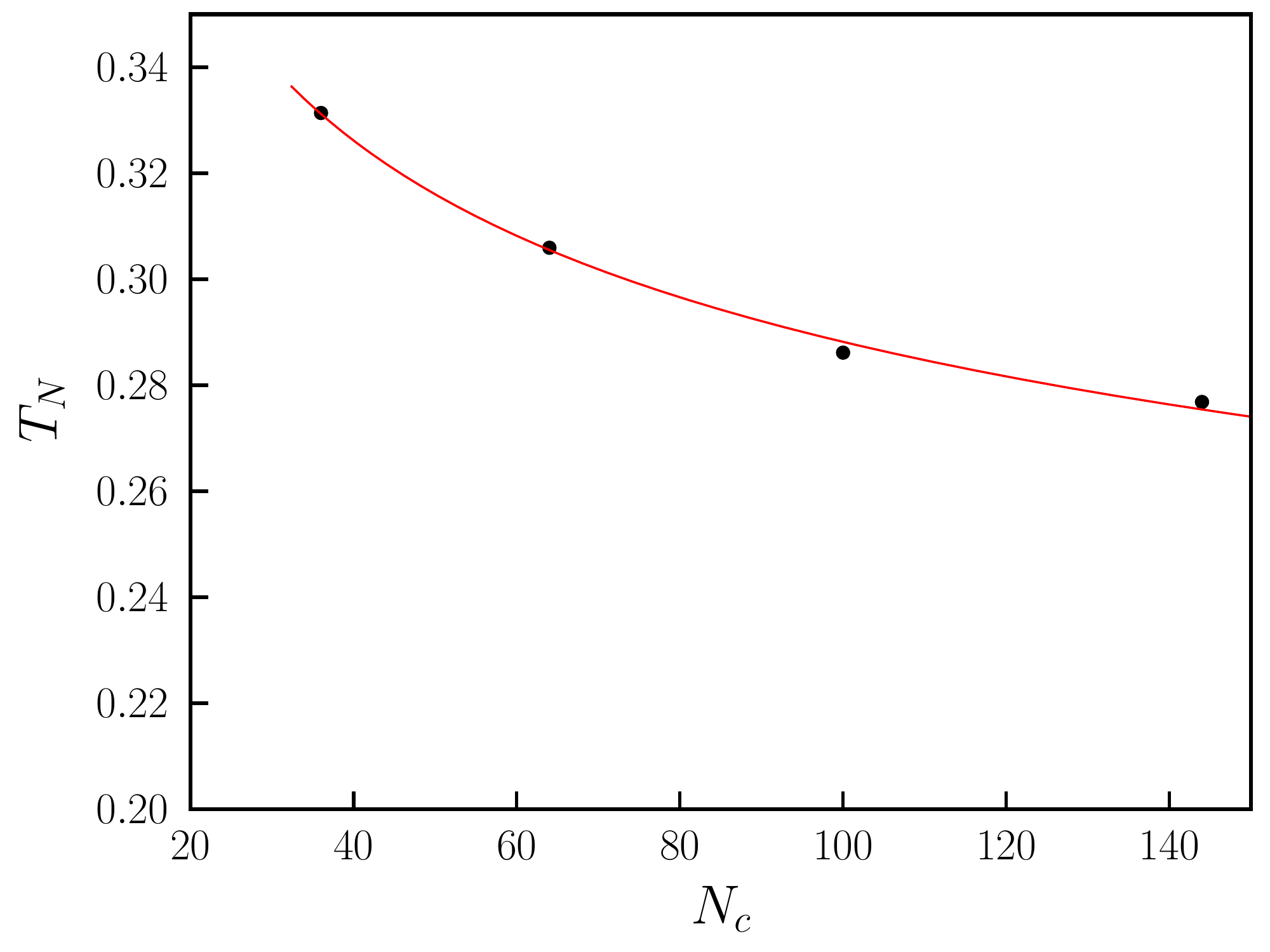} 
	\end{center}
	\caption{\label{fig:Tn_versus_Nc} \dcaplus results for the Ne\'el temperature $T_N$ versus cluster-size for $U/t=7$ at half-filling. The red curve shows the logarithmic decay of the Ne\'el temperature according to Eq.~(\ref{exp_decay}).} 
\end{figure}

We start our investigation of the intermediate coupling regime $U=7t$ by studying magnetism in the half-filled model, which is known to become antiferromagnetic at $T=0$. Mean-field methods such as the DMFT or DCA, however, due to their mean-field character at a finite cluster size, find an antiferromagnetic state at a temperature $T_{\rm N}>0$, which goes to zero for $N_c\rightarrow \infty$ as observed in previous DCA calculations \cite{Maier2005PRL}. This problem therefore provides another interesting test-bed to examine the cluster size dependence of the \dcaplus algorithm. 

In two dimensions, the antiferromagnetic correlation length develops exponentially as the temperature is lowered, i.e. $\xi \sim \alpha\exp(\gamma/T)$. Then, by assuming that a transition occurs when the correlation length becomes equal to the linear cluster $L_c=\sqrt{N_c}$ at $T=T_{\rm N}(N_c)$, one obtains 
\begin{align}
	\label{exp_decay} \sqrt{N_c} \approx L_c \approx \alpha\, e^{\gamma/T_{\rm N}} \rightarrow T_{\rm N}(N_c) \approx \frac{\gamma}{\log(\alpha^{-1}\:\sqrt{N_c})}\,. 
\end{align}

\begin{figure}
	[t] 
	\begin{center}
		\includegraphics[width=0.5
		\textwidth]{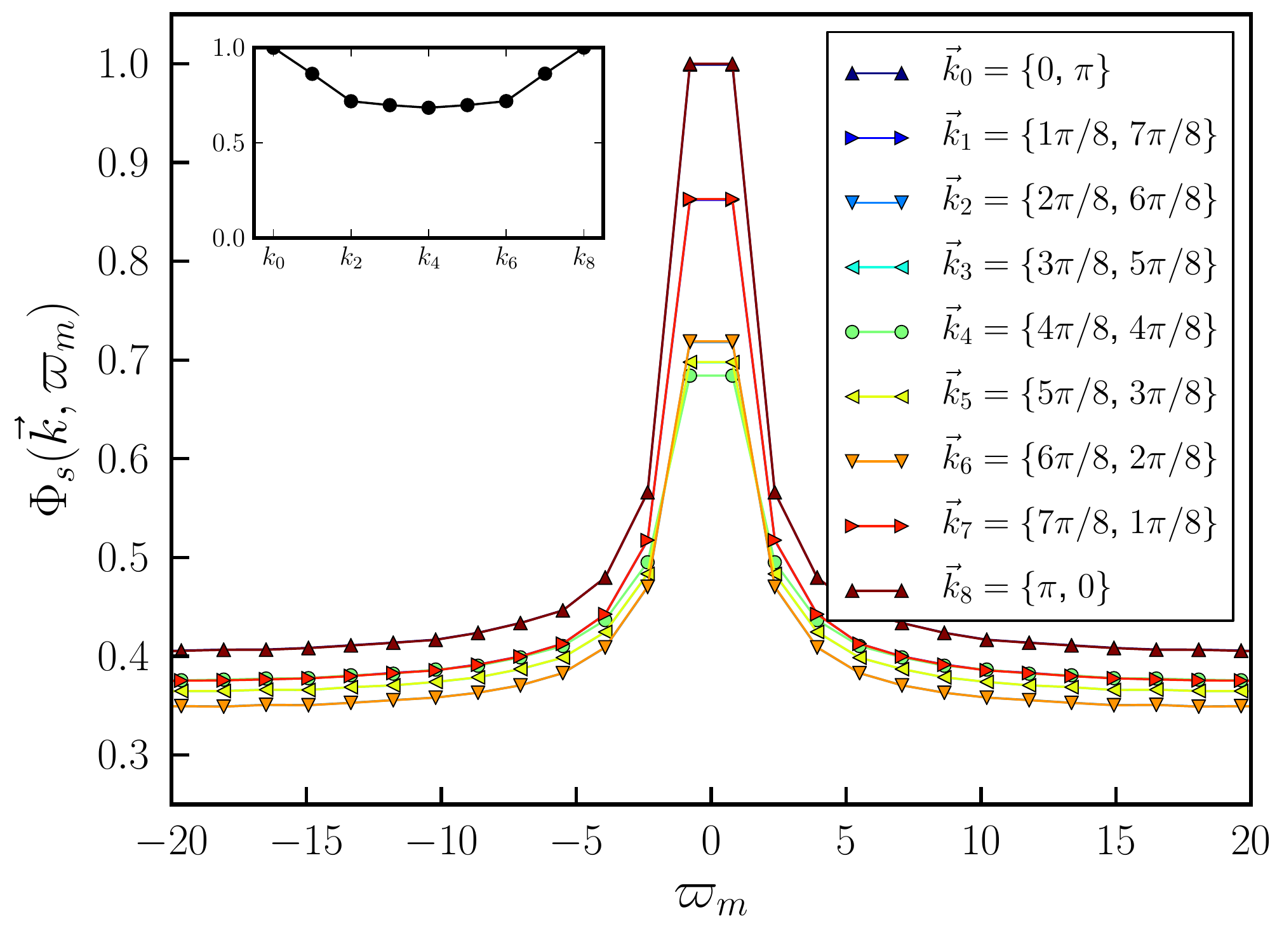} 
	\end{center}
	\caption{\label{fig:Gap_ph} Frequency and momentum dependence of the leading eigenvector in the spin $S=1$ particle-hole channel for $U/t=7$, $N_c=144$ at half-filling for a temperature close to $T_{\rm N}(N_c)$. The inset shows the momentum dependence of $\Phi(k,\pi T)$ along the diagonal from $\vec{k}=(0,\pi)$ to $(\pi,0)$.} 
\end{figure}

Fig.~\ref{fig:Tn_versus_Nc} shows that the \dcaplus results indeed fit this logarithmic decrease of $T_{\rm N}$ with $\sqrt{N_c}$. Here we have determined $T_{\rm{N}}$ from $\lambda(T_{\rm{N}})=1$, where $\lambda$ is the leading eigenvector of the lattice Bethe-Salpeter equation in the spin $S=1$ particle-hole channel for $Q=(\pi,\pi)$. The frequency dependence of the corresponding eigenvector $\Phi(\vec{k},\varpi)$ is shown in Fig.~\ref{fig:Gap_ph} for a selected set of momenta $\vec{k}$. The weak momentum dependence of $\Phi(\vec{k},\varpi)$ indicates that the effective interaction giving rise to the antiferromagnetic state is local. And its frequency dependence reflects a mostly instantaneous interaction which also has a retarded component for this strength of the Coulomb interaction. 

We now turn to the doped model at $U=7t$ and study the superconducting transition for a filling of $\langle n \rangle =0.9$. For these parameters, the standard DCA algorithm can only access clusters as large as 12 sites because of the fermion sign problem. The \dcaplus algorithm, however, significantly delays the sign problem and allow us to access clusters as large as 28 sites. 

\begin{figure}
	[t] 
	\begin{center}
		\includegraphics[width=0.5
		\textwidth]{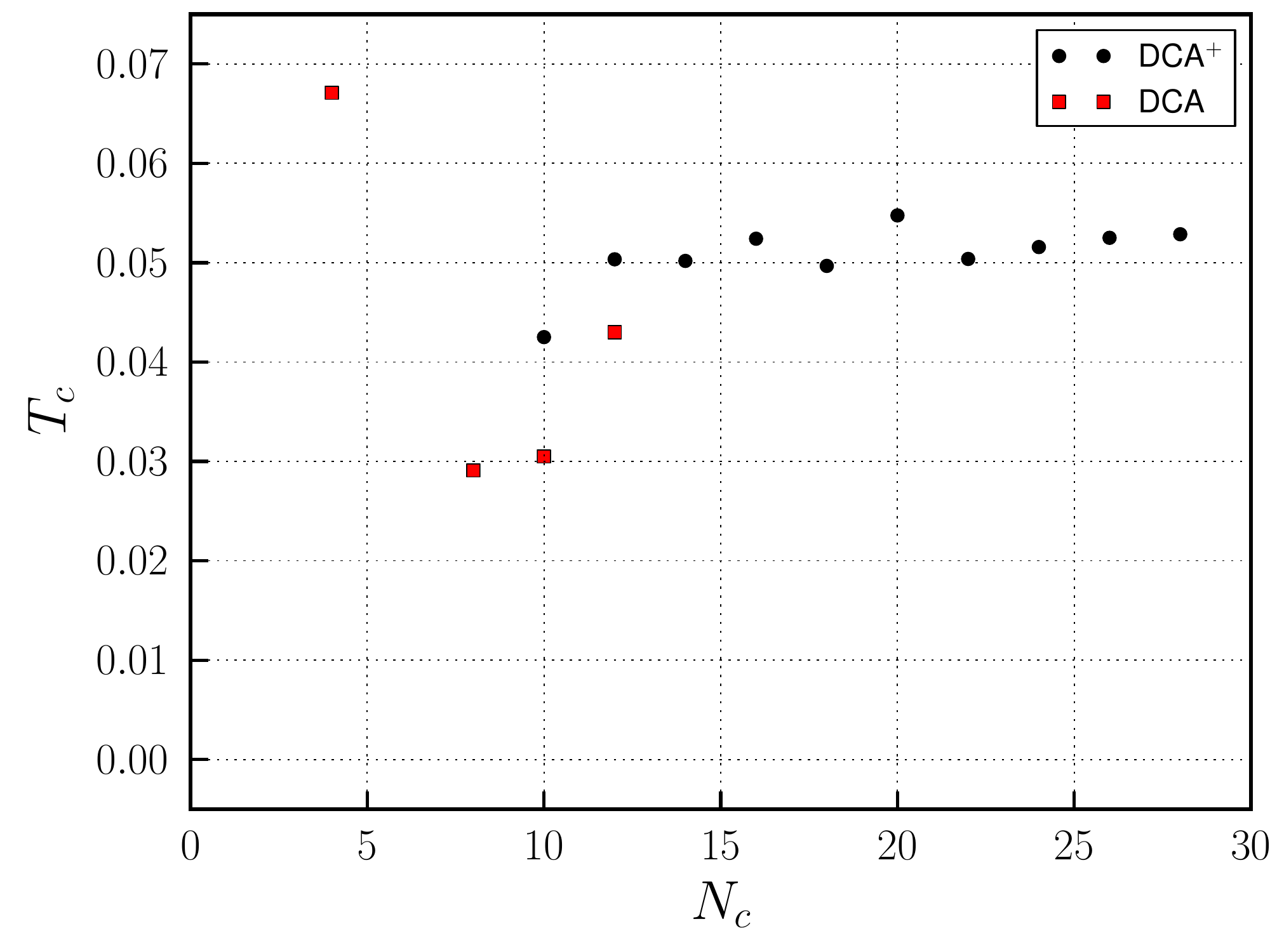} 
	\end{center}
	\caption{\label{fig:Tc_versus_Nc_7} DCA (red squares) and \dcaplus (black circles) results for the superconducting transition temperature $T_c$ versus cluster-size for $U/t=7$ and 10\% doping. } 
\end{figure}

Fig.~\ref{fig:Tc_versus_Nc_7} shows the \dcaplus results for the superconducting transition temperature $T_c$ versus cluster-size (black circles) in addition to the DCA results (red squares). The DCA data for $T_c$ have significant cluster size dependence and irregular behavior and it is impossible to determine an estimate of $T_c$ based on these results. In contrast, the \dcaplus results are much more systematic: Similar to the weak coupling $U/t=4$ case, one observes a small cluster regime in which $T_c$ increases with $N_c$, followed by a regime where $T_c(N_c)$ appears approximately constant. Interestingly, the second regime of constant $T_c$ is reached already for a significantly smaller cluster size than for the weak coupling case. From this we estimate the coherence length $\xi \approx \sqrt{12} \approx 3.5$ lattice spacings for $U=7t$ and $\langle n\rangle =0.9$. This is about half of the estimate we obtained for $U=4t$ and therefore is consistent with the general expectation that the coherence length decreases with increasing interaction strength $U$.

\begin{figure}
	[t] 
	\begin{center}
		\includegraphics[width=0.5
		\textwidth]{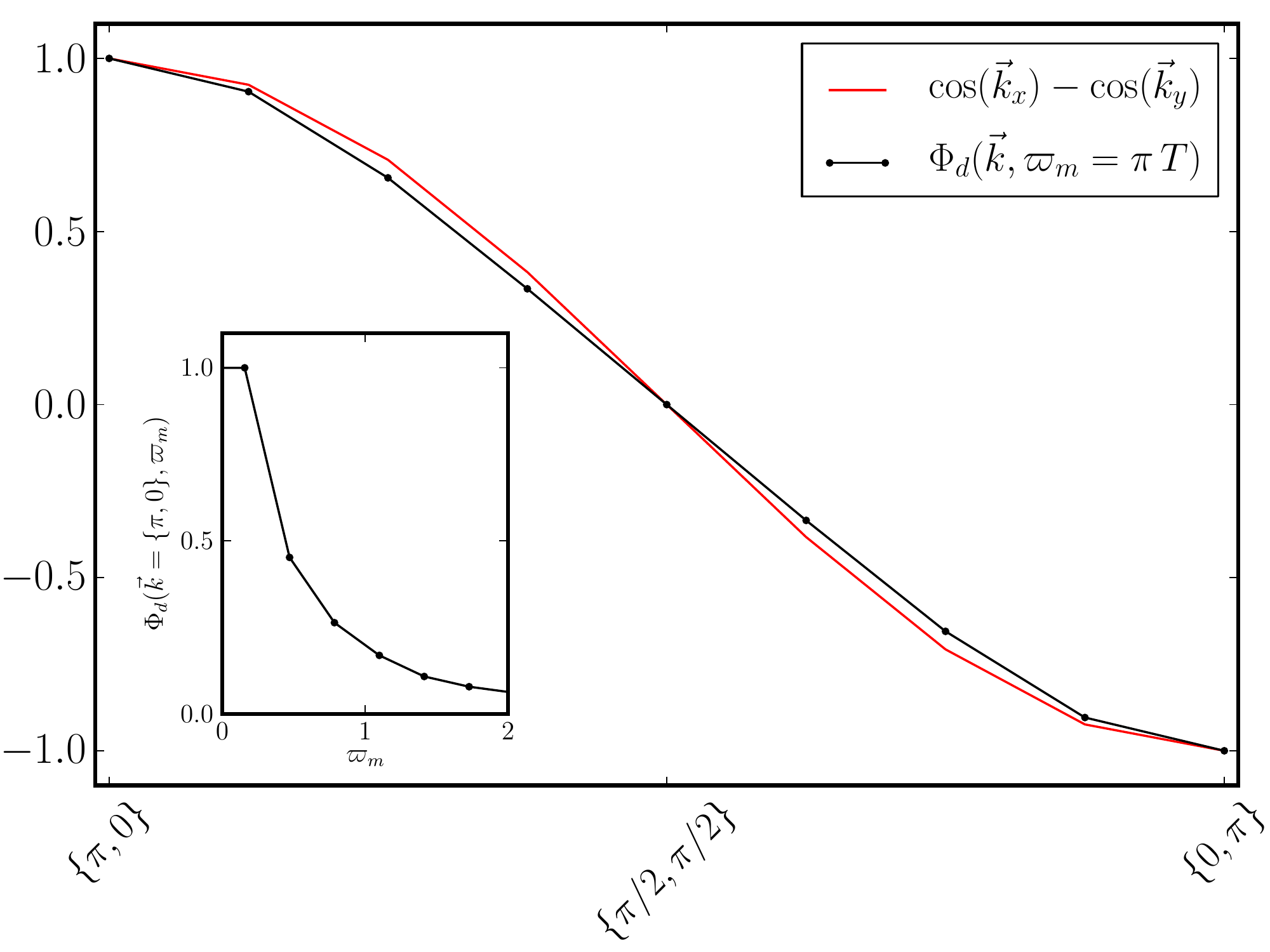} 
	\end{center}
	\caption{\label{fig:Gap_pp} The k-dependence of the leading eigenvector at the first Matsubara frequency in the particle-particle channel for $U/t=7$, $\beta=20$, $N_c=24$ and 10\% doping. One can clearly observe the $d_{x^2-y^2}$ $\cos k_x-\cos k_y$ structure (red-line). Inset: the $\varpi$-dependence of $\Phi(k=\{\pi,0\},\varpi)$.} 
\end{figure}

The $\vec{k}$ dependence of the leading d-wave eigenvector $\Phi({\vec k},\varpi_0=\pi T)$ obtained for the $N_c=28$ site cluster is plotted in Fig.~\ref{fig:Gap_pp}. Its d-wave $\cos k_x - \cos k_y$ structure is obvious from this plot. A detailed analysis of the contribution of higher d-wave harmonics will be published elsewhere. The $\varpi$ dependence of $\Phi({\vec k},\varpi)$ reflects the frequency dependence of the pairing interaction \cite{Maier2006PRB} and is shown for ${\vec k}=(\pi,0)$ in the inset. From this one sees that $\Phi({\vec k},\varpi)$ falls off with $\varpi$ on a scale set by $J=4t^2/U \approx 0.57$. This reflects a retarded pairing interaction with similar dynamics as the spin-fluctuations \cite{Maier2006PRB}. 

\section{Conclusion}

In this paper, we have presented an extension of the recently introduced \dcaplus algorithm to the calculation of two-particle correlation functions. The \dcaplus extends the dynamic cluster approximation with a continuous self-energy and thereby reduces its cluster shape depndencies and the fermion sign problem of the underlying QMC solver. The \dcaplus two-particle framework is derived from the requirement of thermodynamic consistency, which assures that quantities calculated from the two-particle Green's functions are identical to those calculated from the single-particle Green's function. We have shown that this requirement is satisfied if the coarse-grained vertex function $\bar{\Gamma}^\alpha(K,K') = \int d\vec{k}d\vec{k}'\phi_{\vec{K}}(\vec{k})\Gamma^\alpha(k,k')\phi_{\vec{K}'}(\vec{k}')$ is equal to the corresponding vertex function calculated on the cluster, $\Gamma^\alpha_c(K,K')$. This is analogous to the \dcaplus constraint on the single-particle level, which requires the coarse-grained self-energy $\bar{\Sigma}(K) = \int d{\vec{k}} \phi_{\vec K}(\vec{k}) \Sigma(k)$ to be equal to the cluster self-energy $\Sigma_c(K)$. We have then presented a procedure to determine the lattice vertex function $\Gamma^\alpha(k,k')$ from the cluster vertex function $\Gamma^\alpha_c(K,K')$ through inversion of the constraint. This procedure consists of a singular value decomposition of the cluster vertex $\Gamma^\alpha_c(K,K')$, followed by an interpolation of the singular vectors and a subsequent deconvolution of the interpolated cluster vertex. 

We have validated the \dcaplus two-particle framework using the 2D attractive Hubbard model, for which previous large scale finite size QMC results are available. We have determined the $s$-wave superconducting transition temperature $T_{\rm KT}$ in the doped model using two complementary procedures: (1) Using a data-collapse of the $s$-wave cluster pair-field susceptibility and (2) using the lattice irreducible particle-particle vertex computed with the new framework to determine the leading eigenvalue of the Bethe-Salpeter equation. Both methods employed a Kosterlitz-Thouless scaling behavior to determine the exact infinite cluster size result $T_{\rm KT}$ and were shown to give identical results for $T_{\rm KT}$. Moreover, the \dcaplus results were shown to confirm the earlier finite size QMC results.  

We then presented calculations for the 2D repulsive Hubbard model, for both the weak coupling $U=4t$ and intermediate $U=7t$ regimes. For $U=4t$, we have found that the \dcaplus significantly improves upon earlier DCA calculations of the superconducting $d$-wave $T_c$ in the doped model with $\langle n\rangle=0.9$. While the DCA calculations could only access cluster sizes up to 26 sites and gave results with erratic cluster size dependence, the \dcaplus calculations can access cluster sizes up to 56 sites and the cluster size dependence is systematic: For small clusters, $T_c$ increases systematically with cluster size, while for larger clusters it decreases weakly consistent with Kosterlitz-Thouless behavior. By scaling to infinite cluster size we were able to estimate $T_{\rm KT}=0.0199\pm 0.0020$ for $U=4t$. Furthermore, we have argued that the change in the cluster size dependence happens when the linear cluster size becomes of the order of the superconducting coherence length $\xi$. From this we estimate $\xi \sim 6$ lattice spacings for $U=4t$ and $\langle n\rangle=0.9$.

For $U=7t$ and $\langle n\rangle=0.9$, we were able to access clusters up to 28 sites, a significant improvement over the maximum DCA cluster size of only 12 sites. As for the weak coupling regime, the \dcaplus results display systematic behavior as a function of cluster size. For clusters larger than 12 sites, $T_c$ appears to saturate at a value of $T_c \sim 0.053$, i.e. significantly larger than our estimate of $T_c$ for $U=4t$, and from the cluster size dependence we estimate a coherence length $\xi \sim 3.5$ lattice spacings for $U=7t$ and $\langle n\rangle=0.9$. The leading eigenvector of the particle-particle Bethe-Salpeter equation close to $T_c$ is shown to follow a $d_{x^2-y^2}$ $\cos k_x - \cos k_y$ dependence and its frequency dependence indicates a pairing interaction that is retarded on a scale set by the exchange energy $J=4t^2/U$.

In summary, we have shown that the \dcaplus algorithm provides a significant improvement over the DCA approach in the calculation of two-particle properties and the determination of phase instabilities. The reduced fermion sign problem and improved cluster shape and size dependence allows us to access significantly larger clusters at lower temperatures and larger interaction strengths and provides results with systematic cluster size dependence. This enables the reliable extraction of transition temperatures by scaling the results to infinite cluster size and thus facilitates an accurate study of the full temperature versus doping phase diagram of the 2D Hubbard model for realistic parameters relevant to the cuprates. 

\section*{Acknowledgements}
This research was carried out with resources of the Oak Ridge Leadership Computing Facility (OLCF), the Swiss National Supercomputing Center (CSCS), and the Center for Nanophase Materials Sciences (CNMS). OLCF and CNMS are located at Oak Ridge National Laboratory and supported, respectively, by the Office of Science under Contract No. DE-AC05-00OR22725 and by the Scientific User Facilities Division, Office of Basic Energy Sciences, of the Department of Energy.

%
\bibliography{refs_correct.bib}

\end{document}